\documentclass{ws-ijmpa}

\begin{document}

\newcommand{\Tr}{{\rm Tr}}
\newcommand{\lpartial}{\buildrel \leftarrow \over \partial}
\newcommand{\rpartial}{\buildrel \rightarrow \over 
\partial}
\newcommand{\np}{{\rm :}}
\newcommand{\llbracket}{[\hspace{-1.6pt}[}
\newcommand{\rrbracket}{]\hspace{-1.6pt}]}
\newcommand{\llbracketbar}{[\hspace{-1.5pt}[ \hspace{-7pt}\smallsetminus}
\newcommand{\rrbracketbar}{]\hspace{-1.5pt}]\hspace{-10pt}\smallsetminus}
\newcommand{\vertbar}{\hspace{3pt}|\hspace{-9pt}\smallsetminus\hspace{-3pt}}
\newcommand{\slashldelimit}{\hspace{-3pt}\smallsetminus}
\newcommand{\rhdllbra}{\triangleright\hspace{-7pt}[\hspace{-1.8pt}[ \hspace{2.5pt}}
\newcommand{\rhdrrbra}{\, \triangleright \hspace{-5pt}\rrbracket \,}
\markboth{Authors' Names}
{An attempt towards field theory of D0 Branes}

%
\catchline{}{}{}{}{}
%

\title{An Attempt Towards Field Theory of D0 Branes \\
-- Quantum M-Field Theory --
\footnote{
Invited talk presented at the workshop 
``International Conference on Progress of String Theory 
and Quantum Field Theory", Osaka City University, December, 
2007, to be published in the proceedings. 
}}

\author{TAMIAKI YONEYA
}

\address{Institute of Physics, University of Tokyo, Komaba\\
Meguro-ku, Tokyo, 153-8902, Japan
\\
tam@hep1\_c\_u-tokyo\_ac\_jp}



\maketitle


\begin{abstract}
I discuss my recent attempt in search of a new framework for quantum field theory of D branes.  
After explaining 
some motivations in the background of this project, 
I present, as a first step 
towards our goal, a 
 second-quantized reformulation of the U($N$) Yang-Mills quantum 
mechanics in which the D0-brane creation-and-annihilation fields 
connecting theories with different $N$ are introduced. Physical observables are expressed in terms of bilinear forms of the D0 fields. The large $N$ limit
is briefly treated  using this new formalism.

\end{abstract}

\ccode{PACS numbers: 11.25.Hf, 123.1K}

\section{Introduction: motivations}	

 One of the most characteristic features of D-brane dynamics is borne out by the duality between open and closed strings. In particular, the open-closed string duality is clearly at the foundation of gravity/gauge correspondence through D branes. In spite of its paramount importance, however, 
the open-closed string duality has been formulated 
only  in terms of the perturbative world-sheet picture. The current candidates of 
non-perturbative string theory, such as string field theory 
and various versions of matrix models,  thus far have not been 
providing any deeper insight.  

In this talk I would like to present some considerations, 
 on the basis of my 
previous work,\cite{d0field} pointing towards a possible non-perturbative understanding 
of the open-closed duality through the notion of quantum 
fields for D branes.  The essential idea 
can be explained by drawing a simple analogy 
with the so-calld Mandelstam 
duality which holds between 
the sine-Gordon model and the massive Thirring model in 
two-dimensional field theory. On the bulk-gravity side describing their dynamics 
in terms of closed strings, D branes appear 
as lump (or soliton-like) configurations with or without sources. 
 On the other hand, D branes are also treated using open strings 
propagating longitudinally along D branes. 
The latter description, introducing the coordinates of D branes explicitly,  is a `configuration-space' formulation which 
is similar to the ordinary multi-particle quantum mechanics. 
In analogy with the Mandelstam duality, the former corresponds to the 
sine-Gordon description of solitons, while the latter amounts 
to treating the D brane as an 
elementary excitation corresponding to the Dirac fields 
 of the massive Thirring model. 
In the context of 
two-dimensional field theory, these two descriptions are 
related by the bosonization (from the latter to the former) or fermionization (vice versa).  Basically, various observables 
of the sine-Gordon model are represented in terms of 
bilinear forms of the Dirac fields in the Thirring model.  

It seems quite natural to expect that if we were able to reformulate the 
latter configuration-space formulation in a fully second-quantized 
form introducing fields corresponding to D branes, there is a chance to 
establish a similar duality relation between the above two 
descriptions of D branes.  What we are imagining is illustrated with 
the following diagram. The D-brane field theory 
would  hopefully  provide the third possible formulation 
of string theory which could pave a new route 
connecting open and closed strings in a non-perturbative 
fashion. 


\begin{figure}[htbp]
\begin{center}
\includegraphics[width=9cm]{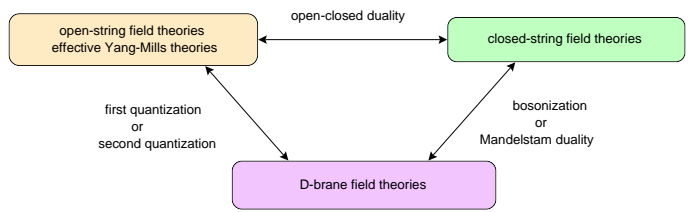}
\vspace{0.1cm}
\caption{\footnotesize{
A trinity of dualities: D-brane field theories point toward 
the third possible formulation of string theory treating 
D-branes as elementary excitations. 
}}
\end{center}
\vspace{-0.3cm}
\end{figure}
As a first step to our goal, we attempt to second-quantize 
the Yang-Mills quantum mechanics, corresponding
to the arrow on 
the left-hand side of this diagram. Namely, 
we reformulate the whole content of the Yang-Mills 
quantum mechanics in the Fock space which 
unites all different sizes of the U($N$) gauge group. 
 As we discuss in the next 
section, such an attempt 
 requires us to enlarge  considerably the usual framework 
of field theory, especially with respect to 
quantum statistics. 

There is another motivation for our attermpt. According to the 
well-known BFSS conjecture,\cite{bfss}  the U($N$) D0 (super) 
matrix quantum mechanics 
may also be interpreted as an {\it exact} formulation of M-theory 
in a special infinite-momentum frame, in which the whole 
system is boosted along the 
compactified direction, 
the tenth spatial dimension being a circle of 
radius $R_{10}=\ell_s g_s$ $\, (g_s \sim g_{{\rm YM}}^2)$, with an infinitely large momentum 
$
P_{10}=N/R \rightarrow \infty$. 
It is then desirable to treat $N$ as a genuine dynamical 
variable rather than as a mere parameter of the theory. 
The large $N$ limit  involved 
here is not the usual  planar 
limit with $g_sN$ begin kept fixed. Also in this aspect, it seems 
 worthwhile
 to pursue an entirely new approach in which 
the Yang-Mills quantum mechanics is formulated in a 
completely unified way for all different $N$.  We can 
hope for example that 
if one could find the description of the same theory in 
general Lorentz frame in eleven dimensional space-time, 
the resulting theory 
would necessarily include anti-D0 branes in addition to 
D0 branes when it is reduced to ten dimensions.

\section{Gauge invariance as the quantum-statistical symmetry}

   From the viewpoint of second-quantization, the Yang-Mills 
quantum mechanics exhibits a peculiar feature with respect to 
quantum-statistical symmetry. In usual $N$-particle quantum 
mechanics in the configuration-space formulation, the 
wave function $\Psi(x^i_1, x^i_2, \ldots, x^i_N)$ involves $N$ (vector) coordinates $x^i_a
\, 
(a=1, 2, \ldots, N)$ and other necessary variables of each 
particle.\footnote{
Throughout this report, the spatial dimensions should be 
understood as $d=9$, but we use 
notation such as $d^d x$  for the integration measure for 
bosonic variables, 
since the formalism is valid for any dimensions, until 
we take into account the supersymmetry. 
The Grassmannian coordinates will be suppressed for simplicity. } Depending on bosons or fermions, 
the wave functions must be totally symmetric or anti-symmetric 
with respect to arbitrary permutation of particles. 
In the Yang-Mills quantum mechanics for D particles, the coordinates of 
particles are replaced by $N\times N$ (vector) matrices 
$X^i_{ab}$, and the wave function $\Psi[X^i_{ab}]$
 must be 
invariant under continuous U($N$) 
transformations, 
\begin{equation}
X^i_{ab}\rightarrow \left(X^i\right)^{U(N)}_{ab}=\left(U(N)X^iU^{-1}(N)\right)_{ab}.
\label{unitarysymmetry}
\end{equation} 
When we set the off-diagonal elements of the matrices to zero, 
the U($N$) transformations are restricted to the 
discrete permutation group S$_N$ as a subgroup of U$(N)$. 
Thus, the role of permutation symmetry in ordinary particle 
quantum mechanics is now played by the 
continuous gauge symmetry. This reflects an essential 
feature of string theory that motion and interaction are 
inextricably connected by its intrinsic symmetry.

In the usual second-quantization in which the above wave function 
appears as a coefficient in superposing base-state vectors in the Fock space as 
\begin{equation}
|\Psi\rangle 
=\frac{1}{N!}\Big(
\prod_{a=1}^N \int d^dx_a\Big) 
\Psi(x_1, x_2, \ldots, x_N)
\psi^{\dagger}(x_N)\psi^{\dagger}(x_{N-1}) \cdots \psi^{\dagger}(x_1)|0\rangle, 
\end{equation}
the validity of the 
canonical commutation relations 
for the field operators crucially depend on the permutation 
symmetry: 
$
[\psi(x), \psi^{\dagger}(y)]=\delta^d(x-y), \quad 
[\psi(x), \psi(y)]=0=[\psi^{\dagger}(x), 
\psi^{\dagger}(y)], 
$
resulting 
\begin{equation}
\psi^{\dagger}
(x_{N})\psi^{\dagger}(x_{N-1}) \cdots \psi^{\dagger}(x_1)|0\rangle
=\psi^{\dagger}(x_{P(N)})\psi^{\dagger}(x_{P(N-1)}) \cdots \psi^{\dagger}(x_{P(1)})|0\rangle
\label{usualstatistics} 
\end{equation}
and 
\begin{equation}
\psi(y)\psi^{\dagger}(x_N)\cdots \psi^{\dagger}(x_1)|0\rangle
={1\over (N-1)!}\sum_P \delta^d(y-x_{P(N)})
\psi^{\dagger}(x_{P(N-1)})
\cdots \psi^{\dagger}(x_{P(1)})|0\rangle
\label{usualannihilation}
\end{equation}
where the summation is over all permutations  $(12\ldots N)\rightarrow 
(a_1a_2\ldots a_N), P(k)=a_{k}$. Here for definiteness we assumed 
the Bose statistics. 

In the case of D particles, in addition to the fact that the permutation symmetry is replaced by 
the continuous unitary symmetry \eqref{unitarysymmetry}, 
 the number 
 of the coordinate-like degrees of freedom 
depends nonlinearly on $N$: the increase of the matrix-degrees of freedom from 
an $N$-particle system to an $(N+1)$-particle system 
is $d(2N+1)=d((N+1)^2-N^2)$.  Evidently, 
these features cannot be 
formulated in the standard canonical 
framework. 
We take this difficulty as a clue 
for exploring a new possible 
language for describing string/M theory, especially the 
aspect of the open-closed string duality, non-perturbatively 
without relying upon the world-sheet picture. 

We now describe our proposal. 
In analogy to the usual second quantization, we first introduce 
`agents', denoted by $\phi^{\pm}[z, \bar{z}]$, which play the role 
of quantum fields creating ($+$) or annihilating ($-$) one D0 brane. 
Here, $(z^i,\bar{z}^i)=\{z_1^i,\bar{z}_1^i, z_2^i, \bar{z}_2^i, \ldots \}$ 
is an infinite-component (complex) spatial vector 
as the base space of our non-relativistic field theory at a fixed time. 
The matrix coordinates are embedded in this infinite-dimensional 
space as follows: the components of the $N\times N$ matrices 
are identified with the components of the complex coordinates, 
suppressing the vector indices,  as
\begin{equation}
z^{(b)}_a\equiv x_a^{(b)}+ i y_a^{(b)}=X_{ab} =\bar{X}_{ba}  \quad \mbox{for}  \quad b\ge a, 
\end{equation}
which is to be interpreted as the $a$-th component 
of the coordinates of the $b$-th D particle.  
An implicit assumption here is that the field algebra, if any,  
should be set up such that we can effectively ignore 
the components 
$z^{(b)}_a$ and $\bar{z}^{(b)}_a$ with $a>b$ for the 
$b$-th operation of adding the D0 branes 
as dummy variables in describing 
the systems with a finite number of D particles. 
Thus in the 4-body case,  for example, 
a $4\times 4$ matrix coordinates are reorganized into 
an array of four complex vectors, 
\[
z^{(1)}=
\begin{pmatrix}
x_1^{(1)}\cr \cdot \cr  \cdot \cr \cdot  \cr
 \cdot \cr 
\cdot \cr
\end{pmatrix},\quad 
z^{(2)}=
\begin{pmatrix}
z_1^{(2)}\cr x_2^{(2)}\cr \cdot \cr \cdot \cr 
 \cdot \cr 
\cdot \cr
\end{pmatrix}, \quad 
z^{(3)}=
\begin{pmatrix}
z_1^{(3)}\cr z_2^{(3)}\cr x_3^{(3)}\cr \cdot \cr
 \cdot \cr 
\cdot \cr
\end{pmatrix}, \quad 
z^{(4)}=
\begin{pmatrix}
z_1^{(4)}\cr z_2^{(4)}\cr z_3^{(4)}\cr x_4^{(4)}\cr \cdot \cr 
\cdot \cr
\end{pmatrix}
\]
  The dots indicate infinitely 
many dummy components.  

The D-particle `fields' creating and annihilating 
a D particle connect the Fock-space states as
\[
\phi^+ : \, |0\rangle \rightarrow \phi^+[z^{(1)}, \overline{z}^{(1)}]|0\rangle 
\, \rightarrow \phi^+[z^{(2)}, \overline{z}^{(2)}]\phi^+[z^{(1)}, 
\overline{z}^{(1)}]|0\rangle \, \rightarrow \cdots ,
\]
\[\hspace{0.75cm}
\phi^- : \, 
0 
\, \leftarrow|0\rangle \, \leftarrow
\phi^+[z^{(1)}, \overline{z}^{(1)}]|0\rangle \leftarrow 
\phi^+[z^{(2)}, \overline{z}^{(2)}]\phi^+[z^{(1)}, 
\overline{z}^{(1)}]|0\rangle \leftarrow \cdots .
\]
The manner in which the degrees of freedom are 
added (or subtracted) is illustrated in Fig. 2. Our 
task is then to formulate these diagrammatic operations  
in terms of appropriate mathematical symbols. 
\begin{figure}[htbp]
\begin{center}
\includegraphics[width=8cm]{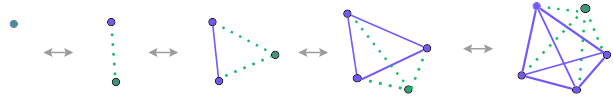}
\caption{\footnotesize{
The D-particle coordinates and the open strings mediating them are 
denoted by blobs and lines connecting them, respectively. 
The real lines are open-string degrees of freedom 
which have been created before the 
latest operation of the creation operator, while the dotted lines 
indicate those created by the last operation.  }}
\end{center}
\end{figure}

\section{Projection conditions and physical operators}
First we have to introduce some new 
structures for taking into account the reduction of 
the components of the base-space coordinates, 
depending on the number of D particles.  
We define the 
following operations of projection
 in the infinite-dimensional base space
\begin{equation}
P_n : \, \, (w_1, w_2, \ldots, w_{n-1}, w_n, w_{n+1}, \ldots) 
\rightarrow (w_1, w_2, \ldots, w_{n-1}, r_n, 0, 0, 0, \ldots),
\label{momentumprojection}
\end{equation}
satisfying $
P_nP_m=P_mP_n=P_m $ for $n\ge m. $ ($w_n=r_n+ i s_n$). 
If the operations are performed in the momentum space, we 
denote them by $\tilde{P}_n$ putting a tilde. Thus for 
arbitrary functions on the base space, 
\begin{equation}
w_k f[P_nw, \overline{P_n w}]=0 \quad \mbox{for}\quad k>n, \quad 
s_n f[P_n w, \overline{P_n w}]=0 \, 
\end{equation}
or 
\begin{equation}
\partial_{w_k} f[\tilde{P}_nw, \overline{\tilde{P}_n w}]=0 \quad \mbox{for}\quad k>n, \quad 
\partial_{s_n} f[\tilde{P}_n w, \overline{\tilde{P}_n w}]=0. 
\end{equation}
The corresponding projection in the Fock space is denoted by 
caligraphic symbols as
\begin{equation}
\phi^+[z, \bar{z}]{\cal P}_n=
{\cal P}_{n+1}\phi^+[\tilde{P}_nz, \tilde{P}_n\bar{z}] , 
\quad 
{\cal P}_n\phi^-[P_nz, P_n\bar{z}]=\phi^-[z, \bar{z}]{\cal P}_{n+1} .
\label{projectionannihilation}
\end{equation}
The vacuum state vector is assumed to satisfy
\begin{equation}
{\cal P}_1|0\rangle =|0\rangle, \, {\cal P}_2|0\rangle=
{\cal P}_3|0\rangle=\cdots = 0 , \quad 
\langle 0|{\cal P}_1 =\langle 0|, \, \langle 0|{\cal P}_2=
\langle 0|{\cal P}_3=\cdots = 0 . 
\label{dualvacuumprojection}
\end{equation}
At first sight, these projection 
conditions may look quite artificial, but should be regarded as the quantum analog of the projector 
property which has been known in the case of 
the lump solutions  in the vacuum string field theory.\cite{vsft}  
A similar feature had previously been observed in the case of 
non-commutative solitons.\cite{noncommutative} In these cases, 
the projector property is exhibited by the {\it  classical} 
solutions. On quantization, such classical 
structures must be reflected into the property of the collective coordinates, constituting the base space for soliton-fields. 

It is convenient and economical  to 
represent the multi-D-particle states 
by indicating only the components of the coordinates on 
which the field operators have nontrivial dependencies, 
suppressing the dummy components. 
We denote the effective $N$-th coordinates 
$ 
(X_{1,N}, X_{N,1}, X_{2, N}, X_{N, 2}, \ldots, 
X_{N-1,N}, X_{N, N-1}, X_{N,N})$ 
by the symbol $X_N$. 
Using this notation, the general base state of $N$-body system is 
expressed as 
\begin{equation}
|N[X]\rangle \equiv {\cal P}_{N+1}\llbracket\phi^+[X_N]\phi^+[X_{N-1}]\cdots 
\phi^+[X_1]\rrbracket|0\rangle,
\label{nbodystate}
\end{equation}
employing a special delimiter symbol $\llbracket *** \rrbracket$ 
for an ordered product ($***$)  of field operators, which 
represents the whole set of the {\it interconnected} blobs and lines for each multi-D0 state in Fig. 2. 

The gauges-statistics symmetry is now expressed  as the 
requirement
\begin{equation}
|N[X]\rangle \simeq |N[X^{U(N)}]\rangle.
\label{gaugestatistics}
\end{equation}
Here the symbol $\simeq$ is used for the purpose of denoting 
that the equality is assumed to be 
valid when operations acting on these 
states are restricted to be `physical'. Physical operations are essentially 
the equivalent of gauge invariant operators in the usual 
first quantized formulation.\footnote{In ref. \cite{d0field},  the operations are classified 
into two classes. Here our discussion is simplified 
by omitting that part. We would like to invite the reader 
to this original paper for a more detailed and precise 
exposition of the technical part of the present approach. 
}

The process of adding one D0 brane is  denoted by using a special symbol as 
\[
|(N+1)[X]\rangle =\phi^+[z](|N[X]\rangle) 
\equiv \phi^+[z] 
\triangleright
|N[X]\rangle 
\]
\[
={\cal P}_{N+2}
\phi^+[X_{N+1}] 
\triangleright\hspace{-7pt}\llbracket \hspace{2.5pt}
 \phi^+[X_N]\phi^+[X_{N-1}]\cdots 
\phi^+[X_1] \rrbracket|0\rangle
\]
\begin{equation}
={\cal P}_{N+2}\llbracket\phi^+[X_{N+1}]\phi^+[X_N]\phi^+[X_{N-1}]\cdots 
\phi^+[X_1]\rrbracket |0\rangle.
\end{equation}
 Even though this is not a gauge invariant operation, it is 
useful as an intermediate tool in formulating physical operations. 

On the other hand, the process of annihilating one 
D0 brane is expressed as 
\[
\phi^-[z,\overline{z}](|N[X]\rangle )
\equiv \phi^-[z,\overline{z}]
{\cal P}_{N+1}\llbracket \phi^+[X_N]
\phi^+[X_{N-1}]\cdots
\phi^+[X_1]\rrbracket
|0\rangle 
\]
\[=\phi^-[z, \bar{z}]{\cal P}_{N+1}\phi^+[X_N]
\rhdllbra 
\phi^+[X_{N-1}]\phi^+[X_{N-2}]\cdots 
\phi^+[X_1] \rrbracket |0\rangle
\]
\[=
\int [dV(N)]\, \delta(x_N-
X^{V(N)}_{NN})\delta^2(z_1-X^{V(N)}_{1N})
\delta^2(z_2-X^{V(N)}_{2N})
\cdots \delta^2(z_{N-1}-X^{V(N)}_{N-1, N})
\sigma_N
\]
\begin{equation}
\times \, {\cal P}_N
\llbracket \phi^+[X^{V(N)}_{N-1}]\phi^+[X^{V(N)}_{N-2}]
\cdots \phi^+[X_{1}^{V(N)}]\rrbracket
|0\rangle . 
\label{Nannihilation3}
\end{equation}
with
$
\sigma_\ell\equiv \delta(y_{\ell})\prod_{k>\ell}\delta^2(z_k) 
$ 
corresponding to the projection condition 
\eqref{projectionannihilation} defined above. 
The integration volume $[dV(N)]$, which is normalized as 
$\int [dV(N)]=N$, 
over the U($N$) transformation $X_a \rightarrow 
X^{V(N)}_a$ is a natural generalization of the 
sum over all possible permutations performed in the 
annihilation process \eqref{usualannihilation} in the usual second-quantization. 

It is to be emphasized that these operations simply {\it 
symbolize} the processes going right or left respectively, 
as illustrated in Fig. 2. 
The multiplication rule of these field operators is {\it maximally} 
non-associative,\cite{d0field} but can be used in order 
to rewrite the Yang-Mills quantum mechanics 
in the second-quantized language. The non-associativity is 
the main reason of introducing new symbols  employed above. 
Conversely, it would not have been possible to require such a 
strong constraint as the 
gauge-statistics condition \eqref{gaugestatistics} unless we abandon the associativity. 

We can also define bra-states by interchanging 
the role of $\phi^+$ and $\phi^-$ and the internal products between bra and ket states, 
leading to the usual probability interpretation for physical 
states. 

As in the usual second-quantization, the whole 
content of configuration-space Yang-Mills quantum mechanics can 
be recast in terms of bilinear forms of the fields, 
\begin{equation}
\langle \phi^+, F\phi^-\rangle 
\equiv 
\int [d^{2d}z]\, \phi^+[z, \bar{z}]\triangleright  F\phi^-[z, \bar{z}] , 
\end{equation}
with $F$ being functions or operators with respect to the 
base-space coordinates acting on the 
D0 fields. For the case of functions 
without derivatives, we find 
\[
\int [d^{2d}z]\, \phi^+[z, \bar{z}]\triangleright  F\phi^-[z, \bar{z}]
|N[X]\rangle
=
\int [dU(N)]\, F[X_{1, N}^{U(N)}, 
X_{2, N}^{U(N)}, \ldots, \bar{X}_{1,N}^{U(N)}, 
\bar{X}_{2,N}^{U(N)}, \]
\begin{equation}
\ldots, X_{N,N}^{U(N)}, 
0, \ldots]|N[X]\rangle, 
\end{equation}
the simplest example of which is the number operator 
with $F=1$, 
\begin{equation}
\langle \phi^+, \phi^-\rangle |N[X]\rangle = N |N[X]\rangle. 
\end{equation}
Another example used below is the potential term $\Tr[X^i, X^j]^2$ appearing in the usual 
Hamiltonian. In terms of the bilinears, it is expressed as
\[
2(\langle \phi^+, \phi^-\rangle+1)
(\langle \phi^+, \Big((\bar{z}^i\cdot z^j)^2-(\bar{z}^i\cdot z^j)
(\bar{z}^j\cdot z^i) \Big)\phi^-\rangle .
\]
Remarkably, the first factor here 
is of order $N$ and hence that 
the second factor can be regarded as being 
well-defined in the usual large $N$ limit. It is 
possible to express arbitrary gauge invariants 
using these bilinear forms, which  almost satisfy 
the standard algebra when acting on physical states.

\section{Schr\"{o}dinger equation and the large $N$ limit}

Using our apparatus, the Schr\"{o}dinger equation turns out to take
 the form
$
{\cal H}|\Psi\rangle =0, 
$ with 
\[\hspace{-7cm}
{\cal H}=i(4\langle \phi^+, \phi^-\rangle +1)\partial_t+
\]
\[\hspace{1.5cm}
2g_s\ell_s\Big(
(\langle \phi^+, \phi^-\rangle +1) \langle \phi^+, 
\partial_{\bar{z}^i}\cdot \partial_{z^i}
\phi^-\rangle 
+3\langle \phi^+, \partial_{\bar{z}^i}\phi^-\rangle\cdot
\langle \phi^+, \partial_{z^i}\phi^-\rangle
\Big)
\]
\begin{equation}
+\frac{1}{2g_s\ell_s^5}
(4\langle \phi^+, \phi^-\rangle +1)(\langle \phi^+, \phi^-\rangle 
+1)\langle \phi^+, \Big(
(\bar{z}^i\cdot z^j)^2 -(\bar{z}^i\cdot z^j)(\bar{z}^j\cdot z^i)
\Big)\phi^-\rangle.
\end{equation}
This is somewhat complicated. But, in the large $N$ limit and in the center-of-mass frame 
satisfying $\langle \phi^+, \partial_{z^i} \phi^-\rangle =0$, 
it is simplified to 
\[
i\partial_t|\Psi\rangle =\Big[
-{g_s\ell_s\over 2}\langle \phi^+, 
\partial_{\bar{z}^i}\cdot \partial_{z^i}
\phi^-\rangle 
-\frac{\langle \phi^+, \phi^-\rangle}{2g_s\ell_s^5}\langle \phi^+, \Big(
(\bar{z}^i\cdot z^j)^2 -(\bar{z}^i\cdot z^j)(\bar{z}^j\cdot z^i)
\Big)\phi^-\rangle\Big]
|\Psi\rangle .
\]
In terms of the M-theory parameters 
$
R=g_s\ell_s\, , \quad \ell_P=g_s^{1/3}\ell_s, \quad 
P^+=\langle \phi^+, \phi^-\rangle/ R
$
after making a rescaling 
$
(z, \bar{z})\rightarrow \left(R/\ell_P\right)^{-1/6}(z, \bar{z})\, , \quad 
t\rightarrow \left(R/\ell_P\right)^{-4/3}x^+/\ell_PP^+, 
$
we find that the Schr\"{o}dinger equation reduces to 
\begin{equation}
P^+\left[P^-+{1 \over 2\ell_P^5}\langle \phi^+, \Big(
(\bar{z}^i\cdot z^j)^2 -(\bar{z}^i\cdot z^j)(\bar{z}^j\cdot z^i)
\Big)\phi^-\rangle\right]|\Psi\rangle=-{1\over 2}\langle \phi^+, 
\partial_{\bar{z}^i}\cdot \partial_{z^i}
\phi^-\rangle |\Psi\rangle, 
\label{finaleq}
\end{equation}
which is essentially bilinear with respect to the D0 fields. 
This expression suggests that the potential term 
of the Schr\"{o}dinger equation in the matrix form may now be interpreted as a light-cone component 
of a special external gauge field 
\[
A^-={1 \over 2\ell_P^5}\left[
(\bar{z}^i\cdot z^j)^2 -(\bar{z}^i\cdot z^j)(\bar{z}^j\cdot z^i)\right]
, \quad A^+=A_{z_n^i} =A_{\bar{z}^i_n}=0.
\]
This particular form, being 
consistent  with supersymmetry,  should appear 
from yet unknown covariant theory through a light-cone 
gauge fixing. 

\section{Conclusion}
What we have discussed in this talk is a first step 
towards a new framework of D-brane field theory. 
There remain innumerable challenging questions. An immediate 
problem would be to study solutions to \eqref{finaleq}. 
Regarding more fundamental questions,  we should explore among 
others (i) possibility of  `transcendental' 
operator algebra  
associated with the continuous statistics and projection conditions; 
(ii)  principle for making the whole 
formalism covariant, especially in eleven dimensional interpretation 
towards quantum `M'-field theory; 
(iii) `bosonization' method through which 
the eleven dimensional supergravity as effective low-energy theory 
or more ambitiously closed-string field theory 
 in ten dimensions would result. 

I would like to thank the organizers for inviting 
me to this wonderful conference. The present work is 
support in part by by Grants-in-Aid for Scientific Research [No. 16340067 (B)] from the Ministry of  Education, Science and Culture of Japan.

\end{document}